Report on the ESO Workshop

# A Synoptic View of the Magellanic Clouds: VMC, Gaia, and Beyond

held at ESO Headquarters, Garching, Germany, 9–13 September 2019

Maria-Rosa L. Cioni[1]
Martino Romaniello[2]
Richard I. Anderson[2]

[1] Leibniz-Institut für Astrophysik Potsdam (AIP), Germany
[2] ESO

The year 2019 marked the quincentenary of the arrival in the southern hemisphere of Ferdinand Magellan, the namesake of the Magellanic Clouds, our nearest example of dwarf galaxies in the early stages of a minor merging event. These galaxies have been firmly established as laboratories for the study of variable stars, stellar evolution, and galaxy interaction, as well as being anchors for the extragalactic distance scale. The goal of this conference was to provide fertile ground for shaping future research related to the Magellanic Clouds by combining state-of-the-art results based on advanced observational programmes with discussions of the highly multiplexed wide-field spectroscopic surveys that will come online in the 2020s.

## Motivations

Observational access to the Magellanic Clouds system was one of the key scientific drivers to build large telescopes in the southern hemisphere, which led to the foundation of ESO itself. Almost 60 years later, the Magellanic Clouds are still very much at the centre of the discourse, providing fundamental insight into several hot research topics. The Magellanic Clouds are our nearest examples of dwarf galaxies at an early stage of a minor merger event. The distribution of their stars and gas provides evidence of an active history of formation and interaction. Thanks to wide and deep photometric observations obtained during the last decade, we have been able to describe the star formation history and the geometry of the Magellanic Clouds at an unprecedented level of detail. The VISTA near-infrared ESO Public Survey of the Magellanic Clouds system (VMC) has played a major role in this endeavour.

In this workshop, the most interesting discoveries emerging from the VMC and other contemporary multi-wavelength surveys were discussed. These results have cemented stellar populations as important diagnostics of galaxy properties. Cepheid stars have, for example, revealed the three-dimensional structure of the system, giant stars have shown significantly extended populations, and blue horizontal branch stars have indicated protuberances and possible streams in the outskirts of the galaxies. The complementary view provided by RR Lyrae stars shows instead regular and ellipsoidal systems. The analysis of the star formation history suggests that the Small Magellanic Cloud (SMC) formed half of its mass prior to about 6 Gyr ago, while a proper motion map reveals tidal features, for example, behind the main body of the galaxy and along the line of sight. The Magellanic Clouds may have arrived at the Milky Way system only recently, together with associated satellite galaxies, and may be more massive than we used to think. The chemical information that was derived, albeit for limited types and numbers of stars, is highly valuable for understanding the internal structure of the galaxies as well as the geometry and chemical evolution of the system — for example, the origin of the Magellanic Stream. For the first time, astronomers are beginning to link age and metallicity distributions with the kinematics and structure of stellar populations, thereby deciphering the formation and evolution of the Magellanic Clouds in great detail.

Within the next year, important observing programmes targeting the Magellanic Clouds will reach completion and provide unique datasets through which the study of the stellar populations will unfold. Moreover, the imminent release of new Gaia data is expected to shed new light on the precision to which stellar populations parameters can be characterised. The Magellanic Clouds remain a unique astrophysical laboratory, for investigations of stellar evolution, star clusters, the distance scale and the measurement of the local value of the Hubble constant to high precision. Future developments focus on using wide-field, high-multiplex spectrographs and powerful images to obtain a robust chemical understanding of the system and use stellar population diagnostics across the Hertzsprung-Russell diagram with unprecedented precision. Most of these developments will culminate in the early 2020s and discussions on how to formulate the most relevant scientific questions are already advanced.

## Summaries of talks and highlights from sessions

The workshop revolved around eight scientific sessions at which a total of 64 talks were presented and a summary is given below. There were also 28 posters complementing each session. A vote was held to select the best among them and the winner was the poster by Raphael Oliveira on "Age and metallicity gradients in the Magellanic Bridge with the VISCACHA survey", which received a prize — a framed photograph of the 30 Doradus press-release image that was produced by the VMC survey.

## The Magellanic Clouds in context

Joss Bland-Hawthorn opened the meeting by highlighting the importance of the Magellanic Clouds as galaxies that have contributed to the growth of the Milky Way. Recent results on the gas distribution, the internal motions of stars and their chemical composition reinforce the Clouds as a place to study many astrophysical processes under different environmental conditions. The orbital history of the Magellanic Clouds, which is reflected in their star formation history, can be used to establish the influence of the Milky Way and to probe the physics of the dark matter halo. Subsequently, Laura Sales reminded us that Lambda-Cold-Dark-Matter ($\Lambda$CDM) substructures around dwarf galaxies indicate that the Large Magellanic Cloud (LMC) must have brought along several of its own dwarf satellites. She argued that, as a result of their recent infall, the dark and baryonic matter would follow a specific path on the sky. Indeed, the combination of deep photometry and accurate astrometry from Gaia has reveleaed that several ultra-faint dwarfs, together with some low-mass classical dwarfs, are consistent with having been accreted as part of the LMC group. This also implies a large LMC virial mass at infall ($M_{200} \geq 3 \times 10^{11}\,M_\odot$) and a direct influence of the LMC on the star formation history of its satellites. Elena Sacchi concluded that the star formation history of ultra-faint dwarfs associated with the Magellanic Clouds differs from that of similar objects associated with the Milky Way.

Satellites that are likely members of the Magellanic Clouds are characterised by bright horizontal branches, dispersed red giant branches and a star formation history that stopped 1–2 Gyr earlier than in ultra-dwarf systems of the Milky Way. Ethan Jahn illustrated, using zoom-in cosmological simulations to study LMC-mass analogues, that tidal interactions with the central galaxy allow the retention of more substructure than in Milky Way-mass hosts, but similarly cause tidal stripping of satellites, suggesting that future kinematical studies will reveal additional satellites associated with the Clouds.

Alice Minelli addressed both the LMC and the Sagittarius dwarf galaxy, the latter being in a more advanced stage of gravitational interaction with the Milky Way than the LMC is. A high-resolution spectroscopic study of 25–30 red giant branch stars per galaxy, measuring the abundance of alpha-, light-, Fe-peak and neutron-capture elements, showed that the two dwarfs experienced a very similar chemical enrichment history despite their current differences, i.e., the LMC still contains gas and presents ongoing star formation while the Sagittarius dwarf is predominantly an old system deprived of gas. Marcel Pawlowski's review dealt with the plane of satellite galaxies problem. Satellite galaxies of the Local Group arrange themselves in narrow structures, with the Magellanic Clouds associated with the Vast Polar Structure, and the Magellanic Stream curiously aligned with other structures, with signs of kinematic correlation (supported by Gaia proper motions) indicative of corotation. Planes of satellites are not common in $\Lambda$CDM simulations and their potential origins include the following: accretions from preferred directions (filaments), group infall or a tidal nature of the dwarf galaxies — there are elements both in favour and against such possibilities.

### Evolution of stars and star clusters in the Magellanic Clouds

The meeting continued with a rich session devoted to the evolution of stars and star clusters, since the Magellanic Clouds provide, in this context, the best samples at sub-solar metallicities. Leo Girardi discussed the calibration of overshooting in main-sequence stars and the evolution of asymptotic giant branch stars, which are critical to determining the nuclear fuel burnt by blue and red stars, and model spectra of 0.1–5 Gyr-old distant galaxies. This work is usually performed in respect of stars that are members of star clusters, where stellar rotation plays a significant role. Field stars, however, where the star formation history is derived, represent a promising input to the calibration of asymptotic giant branch models that also take pulsation and mass-loss into account.

The phenomenon of multiple populations, observed in star clusters of different ages, was reviewed by Nate Bastian. The split main sequence in young (< 1 Gyr old) clusters and the extended main-sequence turnoff in clusters < 2 Gyr old can both be explained using rotation as the dominant mechanism; the large fraction of rapidly rotating stars is supported by a large (~ 60%) fraction of Be stars in these clusters. Seth Gossage demonstrated that stellar evolutionary models that include stellar rotation are able to account for the majority of extendend main-sequence turnoff morphologies. However, other effects like age spreads and braking are not ruled out. Andrea Dupree showed that important constraints to the models of these effects are obtained from high-resolution spectroscopic observations of H$\alpha$ and He I. Furthermore, Ivan Cabrera-Ziri confirmed that neither massive stars nor low-mass stars in young clusters show element abundance variations. On the contrary, a spread in light element abundances (for example, N, Na, C, O, Mg, Al) is likely responsible for the split red giant branches in older clusters. Silvia Martocchia showed the results of a study of about 20 massive (> $10^4\,M_\odot$) star clusters in the Magellanic Clouds where, for the first time, multiple populations were found in clusters as young as 2 Gyr. A larger abundance spread was found in older clusters compared to younger ones, while age differences within a given cluster are < 20 Myr. Complicating the picture, Paul Goudfrooij showed that the faint main sequence of young LMC star cluters is characterised by a kink which is not reproduced by stellar isochrones. This is probably associated with a sudden decrease of temperature resulting from an expansion of the convective envelope in stars with masses < 1.45 $M_\odot$, at the metallicity of the LMC, which may cause braking. Interestingly, the main sequence below the kink of several clusters is consistent with that of a single stellar population.

The properties of the high-mass populations of the Clouds were reviewed by Chris Evans who emphasised the results obtained from the VLT-FLAMES Tarantula Survey. In particular, it was shown that the percentage of binary stars is similar to that in the Galaxy, that it extends to B stars and that there is an excess of massive stars in the region of 30 Doradus with respect to predictions based on the initial mass function. Future observational projects will target SMC massive stars in the ultraviolet to support models at low metallicity. Joachim Bestenlehner focused on the star cluster R136 at the core of 30 Doradus. The cluster age peaks at 1.2 Myr and its most massive stars ($M > 100\,M_\odot$) account for a quarter of the ionising flux and 2/3 of the mechanical feedback. A comprehensive catalogue of 1405 red, 217 yellow and 1369 blue supergiant stars across the SMC was presented by Ming Yang. It stemmed from multi-wavelength observations, from ultraviolet to far-infrared with 29 different filters, and the combination with Gaia data to identify SMC members down to a minimum mass of 6–7 $M_\odot$. Among the intermediate-mass (3–10 $M_\odot$) red supergiants are also Cepheids for which ages strongly depend on model physics — for example, including stellar rotation makes the stars older. Richard Anderson argued that new tests confronting dynamical and evolutionary timescales for Cepheid members of star clusters are needed.

### Star formation history and chemistry across the Magellanic system

Andrew Cole opened the third section of the meeting with a review of the star formation history of the Magellanic Clouds. He highlighted that these nearby galaxies could be both a blessing and a curse, in that they provide an extremely rich population of stars across a Hubble time on the one hand, and on the other hand a level of detail over a large area of sky such that, "no single field is ever going to be totally representative". The LMC had a strong initial phase of star formation that then declined, picking up again 3–5 Gyr ago, while the SMC started forming stars vigorously only 5 Gyr ago.

Star formation histories derived from deep photometry were presented by Tomás Ruiz-Lara from the application of a well-established colour-magnitude diagram fitting technique. Additonal episodes of star formation were identified, as well as differences in the building up of particular regions. The extremes of the LMC bar appear younger and more metal rich than the disc which appears metal poor, but older in the south than in the north. Alessio Mucciarelli reviewed the chemical information obtained from spectroscopic investigations at low resolution (based on the CaII triplet method for a general assessment of the overall metallicity) and at high resolution (based on the abundance of other elements such as Ba and Eu).

Different studies agree on the lower [$\alpha$/Fe] abundance in the LMC compared to that in the Milky Way, which also indicates a lower star formation rate. However, there is disagreement on the slope of [$\alpha$/Fe] as a function of [Fe/H] and on the position of the [$\alpha$/Fe] knee, marking the onset of the influence of type Ia supernovae. Mathieu Van der Swaelmen presented the analysis of FLAMES spectra of red giant branch stars in a few LMC fields. He found that [Mg, O/Fe] are indeed lower in the LMC than in the Milky Way, while [Si, Ca, Ti/Fe] are similar. This suggests that the chemical history of the LMC was dominated by type Ia supernovae and intermediate-mass type II supernovae, without significant differences among the three fields.

Figure 1. Conference photo.

Large-scale infrared surveys have allowed the identification of galaxy-wide samples of young stars, as explained in Joana Oliveira's presentation. This is a crucial step in studying environmental dependencies on star formation and early stellar evolution in order to understand the role of metallicity and galactic structure. In particular, LMC young stellar objects show high accretion rates and significant light-curve variations, while the distributions of upper- and pre-main sequence stars support hierarchical and dust heating substructures. Clifton Johnson showed the powerful impact of the combination of observations with the Atacama Large Millimeter/submillimeter Array (ALMA) and the Hubble Space Telescope (HST) on studies of pre-main sequence stars and their environment in the Magellanic Clouds. The star formation efficiency of molecular clouds in the SMC (~ 2% with 0.5 dex spread) appears consistent with that in the Milky Way and shows a correlation with cloud age.

### Gas and dust within the Magellanic system

Naomi McClure-Griffiths presented an overview of the atomic gas in the Magellanic system, including the Magellanic Bridge, Stream and Leading Arm, which are predominantly gaseous features. A spectacular map of the SMC, with a spatial resolution 10 times higher than that of previous maps, shows a cold (T < 400 K) gas outflow (35–60 km s$^{-1}$), about 40% of which is beyond escape velocity. This implies an HI mass flux of 0.2–1 $M_\odot$ yr$^{-1}$, 2–10 times larger than the rate of star formation in the SMC, which is therefore likely to quench in 0.2–3 Gyr. The rotation curve resembles that of a rotating disc. Both cold and ionised gas (for example, Si I, Si II, H$\alpha$) reside in both the Stream and the Leading Arm. Andrew Fox highlighted the dual chemical origin of the Stream, from both the LMC and the SMC, whilst abundances in the Leading Arm suggest an SMC origin; their variation corroborates a scenario in which the different clumps represent shredded dwarfs accreted as part of the Magellanic group. The average temperature of HI clouds is 30 K and there seems to be a clear correspondence between the location of these clouds and that of the small clumps of CO emission and/or molecular gas, as revelaed by ALMA observations and presented by Katie Jameson. Kat Barger provided an overview of the significant amount of ionising debris surrounding the Magellanic Clouds as revealed by the highest sensitivity emission-line Wisconsin H$\alpha$ Mapper (WHAM) survey. She also showed that supernova explosions in the LMC sustain a large-scale emerging wind (0.4 $M_\odot$ yr$^{-1}$). The level of ionisation in the Stream cannot be explained simply by photoionisation.

The dust content and stellar feedback of the Magellanic Clouds were discussed by Margaret Meixner with a particular focus on the results obtained from projects based on infrared observations with the Spitzer and Herschel space telescopes. The LMC and SMC contain 7.3 × 10$^5$ $M_\odot$ and 8.3 × 10$^4$ $M_\odot$ of dust, respectively, accounted for by asymptotic giant branch stars, red supergiants and supernova production, as well as by dust growth by accretion. The LMC dust is predominantly made of amorphous silicates, while both amorphous silicates and carbon are present in equal amounts in the SMC. Future space missions like the James Webb Space Telescope will allow us to determine the compositon of the dust in the ejecta of supernova 1987A.

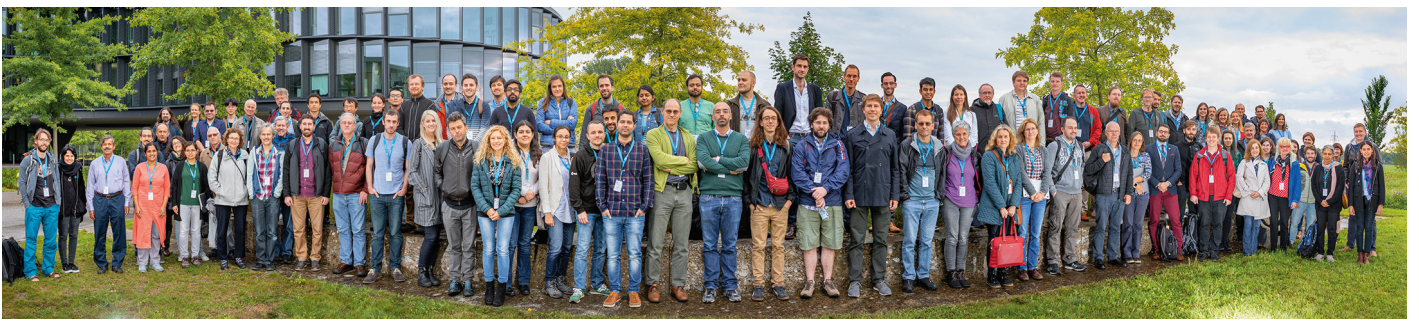

Sikia Gautam studied the correlation between far-ultraviolet (associated with dust) and mid-infrared (associated with polycyclic aromatic hydrocarbon molecules) intensities in many diffuse locations in the SMC, concluding that ultraviolet emission originates in the interstellar field rather than in the intervening medium along the line of sight. Pierre Maggi showed that dust is destroyed as a result of supernova explosions at a rate that depends on the specific element: it is higher for O and Fe and lower for Mg and Si. X-ray survey results also show that the higher numbers of core-collapse supernovae compared to type I in the SMC, compared to the LMC, are perhaps related to their star formation history. Furthermore, supernovae in the LMC appear located in front of the disc (projected onto the bar) or behind it (belonging to 30 Doradus). By studying the spectral energy distribution of background galaxies (from $u$ to $K$ bands and with redshift < 6), Cameron Bell mapped the total instrinsic reddening of the SMC. This method successfully recovers high values in the centre and low values in the external regions utilising galaxies with low levels of intrinsic reddening.

### Internal kinematics and dynamics of the Magellanic Clouds

Denis Erkal demonstrated that the mass of the LMC is large, ~ $10^{11}$ $M_\odot$, and because of that it must have influenced tracers of the Milky Way structure. In particular, a better sky track, distance, proper motion and radial velocity for members of the Orphan stream are obtained when the LMC mass is taken into account. In addition, velocity shifts of Milky Way satellites, warps of the Milky Way disc and a pull of the Milky Way mass internal to 30 kpc can be explained by an LMC influence. Gurtina Belsa argued that only a massive LMC on first infall can maintain the SMC as a binary companion and survive a stable disc after a recent (< 200 Myr) and direct (< 10 kpc) collision. This event is probably responsible for the formation of the one spiral arm and the offset of the disc from the bar in the LMC. At least 30% of Milky Way-type galaxies host an LMC-mass galaxy, with 70% having accreted at least one. Dwarf pairs around these hosts are however rare — they occur in only 6% of cases.

Dana Casetti-Dinescu examined the elemental abundance and three-dimensional kinematics of OB-type stars in the periphery of the LMC, where she confirmed that some stars were born in situ. While the origin of similar stars in the Leading Arm is not clear, they could have formed (and may still be forming) in the LMC disc or in the Milky Way. Lara Cullinane explored the kinematics of the periphery of the LMC using data from the Magellanic Edges Survey (MagES) combined with Gaia. A feature in the northern disc of the LMC shows different kinematics from that of the disc, as obtained from a variety of disc rotation models. Scott Lucchini used hydrodynamical simulations in a tidal scenario to reproduce the Stream and Leading Arm gas.

It is, however, difficult to account for the gas mass in the Stream and the fragmentation of the Leading Arm at the same time. Preliminary simulations including the influence of the Milky Way hot corona show a diffused Leading Arm made predominantly of LMC material and some fragmentation in the trailing arm while matching the observations of the two LMC and SMC filaments. Yang Yanbin showed instead that a hydrodynamical simulation in a ram-pressure-plus-collision scenario is able to reproduce many properties of the Magellanic system, such as the density of HI and the mass of ionised gas in the Stream with a 20% accuracy, the leading arms, and the three-dimensional structure of young and old stars. Florian Niederhofer used data from the VMC survey to calculate the proper motion of young and old stars across the SMC. The median SMC motion and the velocity pattern across tiles are consistent with literature determinations for both samples. Andres del Pino used Gaia data and a neural network method trained on the line-of-sight velocity of young, < 1 Gyr old, stars and on the distance of old RR Lyrae stars, and applied this to > 6 million stars with six-dimensional information, including age and metallicity, to study the distribution and kinematics of stellar populations within the Magellanic Clouds. Their results reproduce two bridges (young and old) separated in distance by about 1.5 kpc, connecting extended distributions characterised by different kinematics.

### The Magellanic Clouds as a distance scale anchor

Grzegorz Pietrzynski led us through the necessary steps to determine an accurate distance to the LMC using eclipsing binaries, one of the primary distance indicators in nearby galaxies, addressing the major sources of error in the calibration of the distance scale (population, extinction, zero-point, blending and physics of the indicator). Many years of extensive photometric and spectroscopic observations were invested to reach an accuracy of 1%.

The role of the LMC in the distance scale was further discussed by Lucas Macri with respect to results on Cepheids and Mira stars, among the secondary distance indicators. In particular, the combination of sparse near-infrared observations and highly sampled optical light curves, as well as single observations with the HST to overcome crowding, provides a significant improvement to the period-luminosity relations used to derive distances, effectively reducing the uncertainty in the Hubble constant. Furthermore, a new periodogram technique based on a multi-band model to fit the light curves will allow us to recover the period of many Mira stars beyond the LMC, to be detected in the future by the Synoptic Survey Telescope at Rubin Observatory. Anupam Bhardwaj showed that period-luminosity relations for Mira stars at maximum light have a 30% less dispersion than at mean light. This is likely due to the destruction of unstable molecules when the Mira is at its warmest phase. Marek Gorski focused on the tip of red giant branch method to derive distances of systems at ~ 2 Mpc from ground-based and ~ 16 Mpc from space-based observations. He highlighted recent improvements to the reddening, edge-detection method and to the near-infrared absolute magnitude calibration.

The most accurate period-luminosity relations for Cepheids in the LMC and SMC, based on near-infrared photometry from the VMC survey, were presented by Vincenzo Ripepi. However, the calibration of these relations, which is based on Cepheids in the Milky Way and the current Gaia data, is still influenced by a metallicity effect and parallax uncertainties. These issues will most likely be fixed in subsequent releases of the Gaia data. To quantify the influence of metallicity on the Cepheid period-luminosity relations, Wolfgang Gieren showed an application of the infrared surface brightness technique to Milky Way, LMC and SMC sources. While the slopes of the relations are not influenced by metallicity, the zero-points are — in the sense that more metal-poor Cepheids are fainter by −0.23 +/− 0.06 mag/dex. Bogumił Pilecki explained that Cepheids in eclipsing binary systems allow us to derive physical parameters (for example, period, mass, and radius) from which to obtain evolution and pulsation models. These results place important constraints on, for example, the projection factor (the ratio between the pulsation velocity of the star and its radial motion), a crucial quantity for the calibration of the infrared surface brightness technique. Roberto Molinaro showed the results of fitting non-linear convective pulsation models to the light and radial velocity curves of a sample of Cepheids in both the LMC and the SMC. Extensive grids of models are built for each individual star to derive structural parameters, distance and reddening, as well as to contruct period-luminosity and period-mass relations for comparison with those derived from observations.

### Morphology and structure of the Magellanic Clouds from different stellar populations

Smitha Subramanian analysed data from the VMC survey and showed evidence for a population of red clump stars ~ 12 kpc in front of the SMC, emerging from a region ~ 2.5 kpc away from the centre and towards the east. This population was probably stripped during the last interaction episode with the LMC 300–400 Myr ago. Michele Cignoni presented data from the STEP (SMC in Time: Evolution of a Prototype interacting late-type galaxy) survey, where a bimodal red clump is also detected and where the bright component dominates the Magellanic Bridge. Furthermore, an analysis of blue-loop (core He burning) stars showed that star formation moved from the northeast of the SMC to the southwest; the age ranges in these regions span 120–200 Myr to 120–60 Myr, respectively, while both ranges are present in the central regions.

Dalal El Youssoufi used the VMC survey data to explore the morphology of the Magellanic Clouds, creating maps with a spatial resolution of 0.13–0.16 kpc at different ages. These maps demonstrate in great detail the history of interaction and evolution of the Magellanic Clouds. Anna Jacyszyn-Dobrzeniecka analysed data from the OGLE-IV survey to characterise the three-dimensional structure of the Magellanic Clouds. The clumpy appearance traced by Cepheids contrasts with the regular distribution traced by RR Lyrae stars in both galaxies. The spatial extension of old stars supports the presence of two halos rather than a bridge connecting the LMC with the SMC. Massimiliano Gatto searched for stellar clusters in the outskirts of the LMC using deep photometric data obtained at the VLT Survey Telescope, from the YMCA (Yes, Magellanic Clouds Again) survey, and found 55 new candidates. Most of these clusters are of intermediate age (1–4 Gyr old) with a peak at 2 Gyr and only a few clusters in the age gap (4–10 Gyr).

Doug Geisler showed that the metallicity distribution of stellar clusters in the SMC is bimodal, with peaks at about [Fe/H] = −0.8 and −1.1 dex, and does not show evidence of a strong gradient, while field red giant branch stars have a unimodal distribution (peaked at [Fe/H] = −1.0 dex) and a negative gradient that reverses to positive beyond 4 degrees from the centre of the galaxy. The age-metallicity relation of the clusters shows a significant dispersion at all ages. In the presentation by Noelia Noël, supporting evidence was given for the disruption of the SMC: the gas appears decoupled from the stars, the kinematics of giant stars shows a lot of debris around a bound core and there are breaks in the low surface-brightness profile of young stars. It is also likely that the total mass of the SMC was much larger than the accepted value. Pushing to low surface brightnesses, Vasily Belokurov reviewed the recent studies of the detection of ultra-faint satellites and their association with the Magellanic system; currently 7% of the Milky Way satellites could be brought in by the LMC.

Satellites might also have been destroyed in the LMC group environment and this process most likely created stellar streams. In the LMC the southern arm appears as the counterpart of the northern arm, while there are many other tentacles that are possibly associated with episodes of earlier interactions between the Clouds. Further insight into the periphery of the Magellanic Clouds was given by Gary da Costa (on behalf of Dougal Mackey). Numerous structural distortions were found within the area covered by the MagES survey (~ 1200 square degrees around the LMC and ~ 200 square degrees around the SMC). The offset of several degrees between intermediate-age and old stars in the SMC might be related to an LMC-SMC encounter > 2 Gyr ago.

Young stars in the Bridge form a chain of diffuse clusters in line with HI observations, supporting a feedback process from supernovae and stellar winds. Camila Navarrete confirmed, using spectroscopic observations of individual stars, that two of the streams previously detected from the distribution of blue horizontal branch stars are indeed kinematically coherent structures. On the other hand, the Pisces overdensity is difficult to associate with Magellanic debris, but may be consistent with the expected Magellanic wake into the Galactic halo. The discovery of a young cluster, associated with the Leading Arm because of its distance, metallicity and radial velocity, was presented by Adrian Price-Whelan. This cluster was found from a search of co-moving blue horizontal branch stars and subsequent follow-up studies; it might have formed as a result of the interaction between the Leading Arm and the Milky Way gas.

### Ongoing and future surveys of the Magellanic Clouds

David Niedever opened the last session by presenting results from two surveys: the Survey of the Magellanic Stellar History (SMASH) and the Magellanic Clouds survey using the Apache Point Observatory Galactic Evolution Experiment (APOGEE).

Their combination will set constraints on the evolution of the galaxies. In particular, deep photometry was used to derive a three-dimensional map of the LMC, to detect a warp and a stellar ring in its disc, and to probe the stellar periphery to 21 degrees from the centre. Extensive spectroscopy was used to derive the (low) star formation efficiency compared to that of the Milky Way — supporting their formation in low-density environments and a first infall scenario. Bruno Dias introduced the VIsible Soar photometry of star Clusters in tApii and Coxi HuguA (VISCACHA) survey aimed at the study of stellar clusters in the Magellanic Clouds. The spatial resolution of this survey is better than that achieved by other ground-based photometric surveys because of the use of adaptive optics, which should improve the derivation of the physical properties of stellar clusters (age, mass, reddening, distance, and structural parameters) from the interpretation of colour-magnitude diagrams.

Maria-Rosa Cioni focused on two surveys using the VISTA telescope: the recently completed near-infrared photometric survey VMC and the planned spectroscopic 1001MC (One Thousand and One Magellanic fields) survey. Highlights from the VMC include: the spatial variation of the distribution of mass in the SMC, developing an elongated shape between 5 and 3 Gyr ago, truncated to the west between 500 and 200 Myr ago; the dominant number of 100 Myr-old Cepheids in the northwest of the SMC at closer distances compared to the majority of 200 Myr-old ones in the centre; and the significant distance modulus variation across the LMC and SMC obtained from the tip of the red giant branch method. The scientific goals, area, type and number of targets (about 0.5 million stars and 0.1 million background galaxies) that the 1001MC plans to observe, as part of the consortium that develops the 4-metre Multi-Object Spectroscopic Telescope (4MOST), were also presented. First results from the Galactic ASKAP (Australian Square Kilometer Array Pathfinder) survey, including the Magellanic Clouds and Bridge, were shown by Nickolas Pingel. In particular, he reported the first detection of a break in the power spectrum of HI, demonstrating that this high spatial and spectral resolution survey will allow us to characterise turbulence (and kinematics) to an unprecedented level. A catalogue of OH masers will also be provided.

In the X-ray domain, Frank Haberl reviewed the status of population studies from the XMM-Newton surveys of the Magellanic Clouds and elaborated on future prospects using the eROSITA instrument on board the Spectrum-Roentgen-Gamma satellite. In particular, he highlighted studies of a large sample (~ 120) of high-mass (Be) X-ray binaries in the SMC, correlated with star formation at 25–60 Myr, where only half of them are pulsars. The expected exposure time of eROSITA across the Magellanic Clouds during the course of the survey, details about the instruments, its performance, and the first light commissioning image were also presented. To conclude this session, Knut Olsen presented the Rubin Observatory Legacy Survey of Space and Time (LSST) which is due to begin in 2023 at Vera C. Rubin Observatory. The proposed science case for observing the Magellanic Clouds makes use of the three main advantages of the telescope: i.e., wide, fast and deep. It addresses a broad range of questions that encompass most of the topics discussed in the meeting so far. It also faces technical challenges, such as solving the problem of separating stars from galaxies in dense stellar fields, extracting photometry for objects in these fields, and defining the footprint and cadence of a multitude of repeated observations.

## Main conclusions and ways forward

The workshop was a great success. It provided a crucial platform for the presentation of a state-of-the-art view of the Magellanic system, comparing and combining results from different teams and projects, and stimulating a discussion that brought us to a better understanding of our neighbouring galaxies and their role as important suppliers of material to the Milky Way halo, demonstrating and quantifying the processes related to galaxy interactions, as well as group accretion in general that may be applicable to more distant systems. The workshop enhanced the impact of the VMC ESO Public Survey in the context of other dedicated and complementary programmes, for example using CTIO telescopes, and the plans formulated for future consortium observations of the Magellanic Clouds using the Multi Object Optical and Near-infrared Spectrograph (MOONS) and 4MOST instruments (see, for example, Cioni et al., 2019).

## Demographics

The 104 participants at the workshop came from 17 different countries. The majority were from the United States of America and Germany, with 20% each, followed by ESO, Italy, Australia and the United Kindgom with 10% each; about 60% of the participants were from ESO Member States. Of the attendees, 35% were female and the Science Organising Committee formulated a scientific programme that reflected this percentage. The selection of contributed talks was made without considering the gender of the applicants while invited talks were selected to include, where possible, female speakers. It is interesting to note that the percentage of female participants matched the percentage of females who delivered review presentations. In addition, the workshop had a good balance of career level and seniority. Each of the eight workshop sessions had three review talks and two talks from students.

### Acknowledgements

A big thank you goes to many people: the other members of the Scientific Organising Committee — Kenji Bekki, Andrew Cole, Elena D'Onghia, Eva Grebel, Vanessa Hill, Rolf-Peter Kudritzki, Jacco van Loon, Naomi McClure-Griffiths, and Igor Soszynski — for their valuable help in preparing an excellent scientific programme; the ESO logistics and catering for a smooth and enjoyable experience; the local organiser committee members, Lisa Löbling and Sara Mancino; and, in particular, Stella Chasiotis-Klingner for her effective and swift management of the meeting. The financial contribution from ESO was also instrumental in facilitating the participation of early career scientists.

### Links

[1] Workshop programme: https://www.eso.org/sci/meetings/2019/magellanic_clouds.html
[2] 30 Doradus VMC image: http://www.eso.org/public/news/eso1033/